# Unified Framework for Hybrid Aleatory and Epistemic Uncertainty Propagation via Decoupled Multi-Probability Density Evolution Method

Yi Luo
Institute for Risk and Reliability, Leibniz University Hannover
Callinstraße 34, Hannover, 30167, Germany
Department of Civil and Environmental Engineering, Rice University
6100 Main Street, Houston, 77005, TX, USA
E-mail address: yi.luo@irz.uni-hannover.de

Meng-Ze Lyu*
Department of Civil & Environmental Engineering, Hong Kong University of Science & Technology
Clear Water Bay, Kowloon, 999077, Hong Kong, China
E-mail address: lyumz@ust.hk

Matteo Broggi
Institute for Risk and Reliability, Leibniz University Hannover
Callinstraße 34, Hannover, 30167, Germany
E-mail address: broggi@irz.uni-hannover.de

Marko Thiele
SCALE GmbH
Levelingstr. 40, Ingolstadt, 85049, Germany
E-mail address: marko.thiele@scale.eu

Vasileios C. Fragkoulis
Department of Civil and Environmental Engineering, University of Liverpool
Liverpool, L69 3GH, UK
E-mail address: vasileios.fragkoulis@liverpool.ac.uk

Michael Beer
Institute for Risk and Reliability, Leibniz University Hannover
Callinstraße 34, Hannover, 30167, Germany
Department of Civil and Environmental Engineering, University of Liverpool
Liverpool, L69 3GH, UK
International Joint Research Center for Resilient Infrastructure & International Joint Research Center for Engineering Reliability and Stochastic Mechanics, Tongji

University
1239 Siping Road, Shanghai, 200092, China
E-mail address: beer@irz.uni-hannover.de

**Abstract:**

This paper presents a unified framework for uncertainty propagation in dynamical systems involving hybrid aleatory and epistemic uncertainties. The framework accommodates precise probabilistic, imprecise probabilistic, and non-probabilistic representations, including the distribution-free probability-box (p-box). A central aspect of the framework involves transforming the original uncertainty inputs into an augmented random space, yielding the primary challenge of determining the conditional probability density function (PDF) of the response quantity of interest given epistemic uncertainty parameters. The recently proposed decoupled multi-probability density evolution method (decoupled M-PDEM) is employed to numerically solve the conditional PDF for complex dynamical systems. Several numerical examples illustrate the applicability, efficiency, and accuracy of the proposed framework. These include a linear single-degree-of-freedom (SDOF) system subject to Gaussian white noise with its natural frequency modeled as a p-box, a 10-DOF hysteretic structure subject to imprecise seismic loads, and a crash box model with mixed random and interval system parameters.

**Keywords:** unified treatment; hybrid aleatory and epistemic uncertainty; distribution-free p-box; decoupled multi-probability density evolution method (decoupled M-PDEM).

# 1 Introduction

Modern engineering systems encounter a variety of complex uncertainties throughout their service life (Ang & Tang 1984, 2006). The performance of engineering structures—from super-tall buildings resisting intense earthquakes (Lyu et al. 2024a), and long-span bridges enduring strong winds (Wang et al. 2024), to nuclear power plant containments under disastrous loads (Lyu et al. 2024b), and offshore platforms subject to wave loads (Luo et al., 2022)—is significantly influenced by inherent randomness (aleatory uncertainty) in material properties, geometric parameters, boundary conditions, and spatiotemporal variations in external excitations (Li & Chen 2009). These aleatory uncertainties are fundamentally irreducible, and their probabilistic characteristics are typically derived from long-term observational data (Lyu et al. 2025a). Concurrently, epistemic uncertainties, arising from incomplete understanding and inadequate control over complex systems (Li 1996, Möller & Beer 2004), also play a crucial role. Examples include limited sensor placements in structural health monitoring (Papadimitriou 2004), theoretical simplifications in constitutive models for novel materials (Soize 2000), and limitations in understanding multi-field coupling mechanisms (Ni & Jiang 2020, Behrendt et al. 2022). These uncertainties inevitably lead to incorporate subjective judgments and idealized assumptions into engineering models (Lind 1983, Beer et al. 2013). Distinguishing reducible epistemic uncertainty from irreducible aleatory uncertainty is critical for enhancing structural safety and economic efficiency in engineering decision-making (Der Kiureghian & Ditlevsen 2009).

In uncertainty quantification (UQ), aleatory and epistemic uncertainties are typically modeled using distinct frameworks (Faes et al. 2021). For aleatory uncertainty, probabilistic models provide a comprehensive mathematical characterization of through probability distribution functions, covariance structures, stochastic processes, or random fields (Soize 2017). However, traditional probabilistic approaches may introduce biases when epistemic uncertainty arises due to data scarcity or incomplete knowledge (Ditlevsen 1982, Moens & Vandepitte 2005). Consequently, non-probabilistic models such as interval analysis (Moore et al. 2009, Gao et al. 2010) and fuzzy sets (Zadeh 1965, Yao & Furuta 1986, Wang & Qiao 1993), imprecise probabilistic models including probability boxes (p-boxes) (Ferson et al. 2003), and evidence theory (Toussi & Yao 1982, Helton et al. 2010, Jiang et al. 2013), and hyper-parameter/hierarchical probabilistic models (Sankararaman & Mahadevan 2013, Tao & Chen 2022, Behrendt et al. 2025), have been developed, forming a multi-level toolbox for uncertainty quantification. For instance, interval methods excel when parameter bounds are known but distribution information is lacking (Jiang et al. 2015), while p-boxes provide conservative analyses under small sample sizes by defining bounds on probability distributions (Bi et al. 2019).

Propagation of hybrid aleatory and epistemic uncertainties is a critical scientific challenge requiring synergistic treatment within a unified framework. Conventional double-loop approaches, involving outer loops for epistemic uncertainties and inner loops for aleatory uncertainties, often suffer from prohibitively high computational costs, particularly for large engineering systems (Xiu et al. 2010, Sankararaman & Mahadevan 2011, Liu et al. 2018). Recent developments aim to decouple this nested computational structure using augmented uncertainty spaces. Examples include the extended Monte Carlo simulation (MCS) method combined with importance sampling (Wei et al. 2014) and its enhancement via high-dimensional model representation (HDMR) (Wei et al. 2019a, b), and generalized polynomial chaos expansions (Wang & Ghanem 2021). However, these methods are unsuitable for distribution-free p-box inputs. Alternatively, Faes et al. (2020, 2021) leveraged operator norms of linear systems to decouple uncertainties by first identifying worst-case epistemic parameters. However, this approach remains limited to linear systems, and extensions to weakly nonlinear systems rely on statistical linearization technique (Ni et al. 2022, Jerez et al. 2024). Within the probability density evolution method (PDEM) framework (Chen &

Li 2007), integration with change of probability measure (COM) (Chen & Wan 2019, Wan et al. 2024) has enabled efficient computation in scenarios involving epistemic uncertainty considered by input probability distribution updates (Wan et al. 2020, 2023, Yang et al. 2025). However, its accuracy diminishes for large variation in input distributions, and it lacks universal compatibility across diverse epistemic uncertainty representations (Jiang et al. 2018).

This study introduces a unified framework for propagating epistemic and hybrid uncertainties, leveraging an augmented random space strategy to handle various epistemic uncertainty models, including distribution-free p-boxes. The proposed augmented space strategy transforms the hybrid uncertainty propagation into a unified problem of conditional probability density function (PDF) solution. Specifically, the decoupled multi-probability density evolution method (decoupled M-PDEM) (Lyu et al. 2024d) is employed to efficiently compute high-dimensional joint and conditional PDFs. The decoupled M-PDEM, based on the random event description of the principle of preservation of probability (Li & Chen 2008), decomposes high-dimensional joint PDF solution into sets of manageable one-dimensional partial differential equations, thus overcoming the "curse of dimensionality" and facilitating applications to complex engineering systems with multiple outputs (Lyu et al. 2024c, 2025b, Wan et al. 2025, Xiang et al. 2025).

The remainder of the paper is structured as follows: Section 2 formulates the mathematical problem of hybrid uncertainty propagation, encompassing random variables, probability boxes, and interval models within a generalized propagation framework. Section 3 rigorously derives the decoupled M-PDEM, and introduced its application strategy to hybrid uncertainty propagation. Numerical experiments on representative dynamical systems validate the effectiveness of the proposed approach in Section 4. Finally, Section 5 concludes with a summary and outlines future research directions.

# 2 Augmented Random Space Treatment for Different Types of Uncertainties

## 2.1 Preliminary discussions

In this works, the uncertain parameters are categorized into the following three types based on the information they convey.

1) Random variables, described by precise probabilistic models, are denoted as $\boldsymbol{\theta}_R$. It is described by a probability distribution with joint cumulative distribution function (CDF) denoted as $F_R(\boldsymbol{\theta}_R)$.

2) Imprecise probability parameters. Here take variables modeled by p-box as a representative, denoted as $\boldsymbol{\theta}_P$. For a scalar p-box variable, denoted as $X$, it is described by a set of distribution functions that adhere to the following constrains (Beer et al. 2013):

$$\underline{F}_X(x) \le F_X(x) \le \overline{F}_X(x), \tag{1}$$

$$\int_{-\infty}^{\infty} x \mathrm{d}F_X(x) \in m, \tag{2}$$

$$\left(\int_{-\infty}^{\infty} x^2 \mathrm{d}F_X(x)\right) - \left(\int_{-\infty}^{\infty} x \mathrm{d}F_X(x)\right)^2 \in v, \tag{3}$$

and

$$F_X(x) \in \mathcal{F}. \tag{4}$$

Eqs. (1)-(4) indicate that the CDF falls within the lower and upper bound $\underline{F}_X(x)$ and $\overline{F}_X(x)$; the mean value of $X$ falls within the interval $m$, and variance is bounded by the interval $v$; the distribution belongs to a prescribed admissible distribution set $\mathcal{F}$. For simplicity, it can be denoted as $\left\langle \overline{F}_X(x), \underline{F}_X(x), m, v, \mathcal{F} \right\rangle$.

3) Non-probabilistic parameters, here considering interval variables, denoted as $\boldsymbol{\theta}_I$.

The system's outputs vary depending on the types and combinations of inputs. Similarly,

the outputs fall into one of the three categories.

1) When the inputs consist solely of random variables $\boldsymbol{\theta}_R$, the outputs are also random variables, aligning target problem of classical stochastic dynamic analysis methods, such as the probability density evolution method (PDEM) (Chen et al. 2016, Chen & Wan 2019, Li & Wang 2023).

2) If the inputs comprise only non-probabilistic parameters ($\boldsymbol{\theta}_I$), the outputs are non-probabilistic parameters as well. Theoretically speaking, the PDEM can address this by treating interval parameters as random variables with bounded distributions. However, using the PDEM in this context offers no advantage over optimization strategies.

3) If the inputs include p-box variables $\boldsymbol{\theta}_P$, or any combinations of the three input types, the outputs will be p-box variables.

This study primarily focuses on the latter scenario, where the outputs are p-box variables. The subsequent discussion will be structured according to the different types of inputs.

## 2.2 Hybrid random and non-probabilistic inputs

First consider systems with hybrid random and interval variables as inputs. The interval input parameters $\boldsymbol{\theta}_I$ has the support $[\boldsymbol{a},\boldsymbol{b}]$. The corresponding output quantity of interest, denoted as $X$, would be a p-box parameter. It can be described by CDF set bounded by the lower CDF $\underline{F}_X(x)$ and upper CDF $\overline{F}_X(x)$, i.e., $\underline{F}_X(x) \le F_X(x) \le \overline{F}_X(x)$. The corresponding lower and upper CDF can be given by

$$\underline{F}_X(x) = \min_{\boldsymbol{\theta}_I \in [\boldsymbol{a},\boldsymbol{b}]} F_{X|\Theta_I}\left(x|\boldsymbol{\theta}_I\right), \tag{5}$$

and

$$\overline{F}_X(x) = \max_{\boldsymbol{\theta}_I \in [\boldsymbol{a},\boldsymbol{b}]} F_{X|\Theta_I}\left(x|\boldsymbol{\theta}_I\right), \tag{6}$$

respectively, where $F_{X|\Theta_I}\left(x|\boldsymbol{\theta}_I\right)$ is the conditional CDF with respect to interval input parameters. More strictly, the p-box output is further restricted to the distributed sets given by

$$\mathcal{F}=\left\langle F_{X|\Theta_I}\left(x|\boldsymbol{\theta}_I\right)\right\rangle,\ \boldsymbol{\theta}_I\in[\boldsymbol{a},\boldsymbol{b}]. \tag{7}$$

Therefore, a complete description of the output uncertainty can be given by the p-box $\langle \bar{F}_X(x), \underline{F}_X(x), \mathcal{F}\rangle$.

For system with combined random and fuzzy inputs, the output would be a fuzzy p-box, and the probability information of output can be similarly obtained by changing the support of the input fuzzy variables.

## 2.3 P-box inputs

Then, consider the p-box types of input $\boldsymbol{\theta}_P$, $F_p\left(\boldsymbol{\theta}_p\right)\in\left\langle \bar{F}_p\left(\boldsymbol{\theta}_p\right),\underline{F}_p\left(\boldsymbol{\theta}_p\right),m,v,\mathcal{F}\right\rangle$. No matter distribution-free or parametric, the upper and lower bound of response CDF can all be given by

$$\bar{F}_X(x)=\max\left\{\int_{-\infty}^{x}\int_{-\infty}^{\infty}f_{X|\Theta_p}\left(x|\boldsymbol{\theta}_p\right)f_p\left(\boldsymbol{\theta}_p\right)\,\mathrm{d}\boldsymbol{\theta}_p\mathrm{d}x\right\}, \tag{8}$$

and

$$\underline{F}_X(x)=\min\left\{\int_{-\infty}^{x}\int_{-\infty}^{\infty}f_{X|\Theta_p}\left(x|\boldsymbol{\theta}_p\right)f_p\left(\boldsymbol{\theta}_p\right)\,\mathrm{d}\boldsymbol{\theta}_p\mathrm{d}x\right\}. \tag{9}$$

In these equations, $f_p\left(\boldsymbol{\theta}_p\right)$ is any possible PDF of $\boldsymbol{\theta}_P$. Notice that if only p-box variables are involved in the uncertain inputs, the conditional PDF is a Dirac-$\delta$ function as given by

$$f_{X|\Theta_p}\left(x|\boldsymbol{\theta}_p\right)=\delta\left[x-X\left(\boldsymbol{\Theta}_p\right)\Big|_{\boldsymbol{\Theta}_p=\boldsymbol{\theta}_p}\right]. \tag{10}$$

If random variables $\boldsymbol{\theta}_R$ are also involved in the input, the conditional distribution will spread out to a "distribution".

## 2.4 General discussion

For all the other combinations of the previously discussed types of uncertain parameters,

the problem can be solved by a simple combination of the above-mentioned procedures. Denote all the epistemic uncertainty parameters involved in the system as $\boldsymbol{\theta}_E$. It can be seen that the most crucial step is to obtain the conditional CDF $F_{X|\boldsymbol{\Theta}_E}(x|\boldsymbol{\theta}_E)$ or the conditional PDF $f_{X|\boldsymbol{\Theta}_E}(x|\boldsymbol{\theta}_E)$. Normally, a double loop scheme need be applied to obtain it. Whereas, by assigning a pseudo distribution to the epistemic parameters, which can be represented by the PDF function $\tilde{f}_{\boldsymbol{\Theta}_E}(\boldsymbol{\theta}_E)$, the conditional distribution can be further obtained by

$$f_{X|\boldsymbol{\Theta}_E}(x|\boldsymbol{\theta}_E) = \frac{\tilde{f}_{X\boldsymbol{\Theta}_E}(x,\boldsymbol{\theta}_E)}{\tilde{f}_{\boldsymbol{\Theta}_E}(\boldsymbol{\theta}_E)}. \tag{11}$$

In this equation, the pseudo joint PDF $\tilde{f}_{X\boldsymbol{\Theta}_E}(x,\boldsymbol{\theta}_E)$ is probable to be solved via single-loop schemes. In the next section, the recently developed decoupled M-PDEM (Lyu et al. 2024d) will be implemented.

# 3 Solution via Decoupled M-PDEM

In this section, the theoretical foundation and the numerical implementation of the decoupled M-PDEM (Lyu et al. 2024d) will be introduced. It is developed to efficiently solve the joint PDF of the responses of dynamical systems. However, in the current status, the systems considered involves solely aleatory uncertainty, i.e., random variables, as input.

## 3.1 Theoretical foundation

Consider an $n$-dimensional dynamical system given by the equation of motion as

$$\dot{\boldsymbol{X}}(t) = \boldsymbol{g}[\boldsymbol{X}(t), \boldsymbol{\Theta}, t], \tag{12}$$

where $\boldsymbol{X}(t)$ is the $n$-dimensional response process, and $\dot{\boldsymbol{X}}(t)$ is its time-derivative; $\boldsymbol{\Theta}$ denotes the random source of the system and is an $s$-dimensional random variable; $\boldsymbol{g}(\cdot)$ is an $n$-dimensional deterministic function vector.

Suppose $\boldsymbol{Z}(t)$ is the $m$-dimensional response quantity of interest. The generalized density evolution equation (GDEE), or named as the Li-Chen equation (Nielsen et al. 2016), of the joint PDF of $\boldsymbol{Z}(t)$ and $\boldsymbol{\Theta}$, denoted as $p_{\boldsymbol{Z}\boldsymbol{\Theta}}(\boldsymbol{z},\boldsymbol{\theta},t)$, can be given by (Li & Chen 2008, Chen & Li 2009)

$$\frac{\partial p_{\boldsymbol{Z}\boldsymbol{\Theta}}(\boldsymbol{z},\boldsymbol{\theta},t)}{\partial t}+\sum_{i=1}^{m}\dot{h}_i(\boldsymbol{\theta},t)\frac{\partial p_{\boldsymbol{Z}\boldsymbol{\Theta}}(\boldsymbol{z},\boldsymbol{\theta},t)}{\partial z_i}=0. \tag{13}$$

where $h_i(\boldsymbol{\theta},t)$ is the component of the solution of $\boldsymbol{Z}(t)=\boldsymbol{h}(\boldsymbol{\theta},t)=\left[h_1(\boldsymbol{\theta},t),\cdots,h_m(\boldsymbol{\theta},t)\right]^T$, and $\dot{h}_i(\boldsymbol{\theta},t)$ is its derivative with respect to time. The formal analytical solution of Eq. (13) can be given by

$$p_{\boldsymbol{Z}\boldsymbol{\Theta}}(\boldsymbol{z},\boldsymbol{\theta},t)=p_{\boldsymbol{\Theta}}(\boldsymbol{\theta})\prod_{i=1}^{m}\delta\left[z_i-h_i(\boldsymbol{\theta},t)\right] \tag{14}$$

Notice that if only one component of $\boldsymbol{Z}(t)$, i.e., $Z_i(t)$, is considered, its corresponding Li-Chen equation can be given as

$$\frac{\partial p_{Z_i\boldsymbol{\Theta}}(z_i,\boldsymbol{\theta},t)}{\partial t}+\dot{h}_i(\boldsymbol{\theta},t)\frac{\partial p_{Z_i\boldsymbol{\Theta}}(z_i,\boldsymbol{\theta},t)}{\partial z_i}=0. \tag{15}$$

The analytical solution is

$$p_{Z_i\boldsymbol{\Theta}}(z_i,\boldsymbol{\theta},t)=p_{\boldsymbol{\Theta}}(\boldsymbol{\theta})\delta\left[z_i-h_i(\boldsymbol{\theta},t)\right]. \tag{16}$$

Comparing Eq. (14) with Eq. (16), one has (Lyu et al. 2024d)

$$p_{\boldsymbol{Z}\boldsymbol{\Theta}}(\boldsymbol{z},\boldsymbol{\theta},t)=\frac{1}{\left[p_{\boldsymbol{\Theta}}(\boldsymbol{\theta})\right]^{m-1}}\prod_{i=1}^{m}p_{Z_i\boldsymbol{\Theta}}(z_i,\boldsymbol{\theta},t). \tag{17}$$

Eq. (17) indicates that the $m$-dimensional Li-Chen equation given by Eq. (13) can be analytically treated by solving $m$ decoupled one-dimensional Li-Chen equation given by Eq. (15), which is the key concept of decoupled M-PDEM. Finally, the joint PDF of $\boldsymbol{Z}(t)$, denoted as $p_{\boldsymbol{Z}}(\boldsymbol{z},t)$, can be given by

$$p_{\boldsymbol{Z}}(\boldsymbol{z},t)=\int_{\mathbb{R}^s}p_{\boldsymbol{Z}\boldsymbol{\Theta}}(\boldsymbol{z},\boldsymbol{\theta},t)\mathrm{d}\boldsymbol{\theta}. \tag{18}$$

## 3.2 Numerical implementation

The numerical implementation of decoupled M-PDEM follows the basic steps of numerical realization of the PDEM, but with an additional step in between to assemble the marginal PDF into joint PDF. Specifically, it follows five steps.

**Step 1.** Partitioning the probability space of random input. This further includes the following processes. (1) Generating $n_{\text{sel}}$ representative samples of input variables $\boldsymbol{\Theta}$, denoted as $\boldsymbol{\theta}^{(q)}$, $q=1,2,\cdots,n_{\text{sel}}$. (2) Partitioning the assigned probability domain $\Omega^{(q)}$ of each point $\boldsymbol{\theta}^{(q)}$ based on the Voronoi diagram, and calculating the assigned probability, which is given by

$$P^{(q)}=\int_{\Omega^{(q)}} p_{\boldsymbol{\Theta}}(\boldsymbol{\theta})\mathrm{d}\boldsymbol{\theta}. \tag{19}$$

Numerically, the integration can be computed by MCS. (3) Reformulation of the point set based on generalized F-discrepancy (Chen & Zhang 2013), obtain the final value of $\boldsymbol{\theta}^{(q)}$.

**Step 2.** Representative deterministic analysis. That is, for each representative point $\boldsymbol{\theta}^{(q)}$, solve Eq. (12), and obtain $\boldsymbol{Z}^{(q)}(t)=\boldsymbol{h}\left(\boldsymbol{\theta}^{(q)},t\right)$.

**Step 3.** Solving the Li-Chen equation. For each dimension of $\boldsymbol{Z}(t)$ and for each $\boldsymbol{\theta}^{(q)}$, substitute $\dot{h}_i(\boldsymbol{\theta},t)\approx\dot{h}_i\left(\boldsymbol{\theta}^{(q)},t\right)$ into the corresponding Li-Chen equation [Eq. (15)] and integral on $\boldsymbol{\theta}$ within the subdomain $\Omega^{(q)}$. It yields

$$\frac{\partial p_{Z_i}^{(q)}(z_i,t)}{\partial t}+\dot{h}_i\left(\boldsymbol{\theta}^{(q)},t\right)\frac{\partial p_{Z_i}^{(q)}(z_i,t)}{\partial z_i}=0,\ \text{for } i=1,\cdots,m,\ \text{and } q=1,\cdots,n_{\text{sel}} \tag{20}$$

where

$$p_{Z_i}^{(q)}(z_i,t)=\int_{\Omega^{(q)}} p_{Z_i\boldsymbol{\Theta}}(z_i,\boldsymbol{\theta},t)p_{\boldsymbol{\Theta}}(\boldsymbol{\theta})\mathrm{d}\boldsymbol{\theta}\approx p_{Z_i\boldsymbol{\Theta}}\left(z_i,t\middle|\boldsymbol{\theta}^{(q)}\right)P^{(q)}. \tag{21}$$

The initial condition for Eq. (20) can be cast as

$$p_{Z_i}^{(q)}(z_i,t)=\delta\left[z_i-Z_i^{(q)}(0)\right]P^{(q)}. \tag{22}$$

Eq. (20) can be solved by various numerical schemes, such as the finite difference scheme (Li & Chen 2004) or the mesh free scheme (Wang et al. 2021). Herein, the path integration is used for the solution.

**Step 4.** Assembling the joint PDF for each representative point. The sub-joint-PDF of $\boldsymbol{Z}(t)$ corresponding to the probability sub-domain $\Omega^{(q)}$ can be computed as

$$p_{\boldsymbol{Z}}^{(q)}(\boldsymbol{z},t)=\frac{1}{\left[P^{(q)}\right]^{m-1}}\prod_{i=1}^{m}p_{Z_i}^{(q)}(z_i,t)\text{, for } q=1,\cdots,n_{\text{sel}}. \tag{23}$$

**Step 5**. Integrating the joint PDF. $p_{\boldsymbol{Z}}(\boldsymbol{z},t)$ is calculated by

$$p_{\boldsymbol{Z}}(\boldsymbol{z},t)=\sum_{i=1}^{n_{\text{sel}}}p_{\boldsymbol{Z}}^{(q)}(\boldsymbol{z},t). \tag{24}$$

It should be noted that another significant advantage of the decoupled M-PDEM scheme lies in the storage of high-dimensional joint PDF. For an $m$-dimensional random variable, suppose each dimension is discretized into $n_i$ grid points, where $i=1,\cdots,m$. Then, size of the joint PDF would be $\prod_{i=1}^{m}n_i$. However, via the decoupled M-PDEM scheme, the probability information of the joint PDF is distributed into each representative point with each dimension decoupled. Consequently, the size for storing it would be $n_{\text{sel}}\sum_{i=1}^{m}n_i$. With the increasing of $m$, $n_{\text{sel}}\sum_{i=1}^{m}n_i \ll \prod_{i=1}^{m}n_i$.

The decoupled M-PDEM can be further collaborated with any variance reduction (Bittner et al. 2024) and adaptive sampling schemes (Zhou et al. 2019), similarly as for PDEM, to improve its efficiency and accuracy. However, it is out of the scope of this work, and is not discussed herein.

### 3.3 Decoupled M-PDEM for epistemic/hybrid uncertainty propagation

In order to get the solution of $\tilde{f}_{X\boldsymbol{\Theta}_E}(x,\boldsymbol{\theta}_E)$ in Eq. (11) via the decoupled M-PDEM, one can create a response process vector as

$$\boldsymbol{Z}(t)=\left[X(t),\tilde{\boldsymbol{\Theta}}_E(t)\right]^{\mathrm{T}},\ t\in\left[0,\bar{T}\right] \tag{25}$$

where $\bar{T}$ is the duration of simulated response process $X(t)$; $\tilde{\mathbf{\Theta}}_E(t)$ is the pseudo stochastic process associated with the epistemic uncertainty input parameter, and can be simply defined as a linear process as

$$\tilde{\mathbf{\Theta}}_E(t)=\frac{t}{T}\mathbf{\Theta}_E,\ t\in\left[0,\bar{T}\right]. \tag{26}$$

By solving the joint PDF of $\boldsymbol{Z}(t)$, we could obtain that $p_{\boldsymbol{Z}}(\boldsymbol{z},t)=\tilde{f}_{X\mathbf{\Theta}_E}\left(x(T),\boldsymbol{\theta}_E\right)$. Substitute it into Eq. (11), the condition PDF $f_{X|\mathbf{\Theta}_E}\left(x(T)\middle|\boldsymbol{\theta}_E\right)$ can be obtained.

# 4 Numerical Examples

## 4.1 SDOF linear systems subject to Gaussian white noise

Consider a single-degree-of-freedom (SDOF) oscillator given by the equation of motion as

$$\ddot{X}(t)+2\zeta\omega\dot{X}(t)+\omega^2X(t)=\xi(t), \tag{27}$$

where $X(t)$, $\dot{X}(t)$, and $\ddot{X}(t)$ denotes the displacement, velocity, and acceleration of the oscillator, respectively; $\zeta$ denotes the damping ratio; $\omega$ denotes the natural frequency; $\xi(t)$ is a Gaussian white noise that satisfies $\mathrm{E}\left[\xi(t)\right]=0,\mathrm{E}\left[\xi(t)\xi(t+\tau)\right]=2\pi S_0\delta(\tau)$. In this system, the natural frequency $\omega$ is considered as a p-box variable, defined by $\omega\sim\left\langle\bar{F}_\Omega(\omega),\underline{F}_\Omega(\omega)\right\rangle,\ \omega>0$. Two response statistics are considered as outputs.

The first is the steady-state variance of $X(t)$, denoted as $Y_1$, which is quantified by the p-box $Y_1\sim\left\langle\bar{F}_{Y_1}(y_1),\underline{F}_{Y_1}(y_1)\right\rangle$. Given $\Omega=\omega$, its analytical solution is $y_1=\dfrac{\pi S_0}{2\zeta\omega^3}$. Subsequently, since the determination of $Y_1$ only involves p-box input, according to Eq.

(10), one has $f_{Y_1|\Omega}(y_1|\omega)=\delta\left(y_1-\frac{\pi S_0}{2\zeta\omega^3}\right)$, and $F_{Y_1}(y_1)=1-F_\Omega(\omega)\Big|_{\omega=\left(\frac{\pi S_0}{2\zeta y_1}\right)^{1/3}}$. Further, due to the monotonicity of $Y_1$ with respect to $\omega$, based on Eqs. (8) and (9), the upper and lower bounds of the CDF of $Y_1$ can be analytically given as

$$\begin{cases}\bar{F}_{Y_1}(y_1)=1-\underline{F}_\Omega\left[\omega(y_1)\right]\\ \underline{F}_{Y_1}(y_1)=1-\bar{F}_\Omega\left[\omega(y_1)\right]\end{cases}. \tag{28}$$

The second is the steady-state of $X(t)$. It is denoted as $Y_2$, and quantified by the p-box $Y_2\sim\left\langle\bar{F}_{Y_2}(y_1),\underline{F}_{Y_2}(y_1)\right\rangle$. The uncertainty quantification of $Y_2$ involves both random variables (Gaussian white noise) and p-box variable as inputs. If $\omega$ is given, the steady state of $X(t)$ follows the Gaussian distribution, with zero-mean, and variance equals to $Y_1$. That is,

$$f_{Y_2|\Omega}(y_2|\omega)=\frac{1}{\sqrt{2\pi y_1}}\exp\left[-\frac{y_2^2}{2y_1}\right],\ y_1=\frac{\pi S_0}{2\zeta\omega^3}. \tag{29}$$

Further,

$$F_{Y_2}(y_2)=\int_{-\infty}^{y_2}\int_0^\infty\phi\left(\frac{\tilde{y}_2}{\sqrt{y_1}}\right)f_{Y_1}(y_1)\mathrm{d}y_1\mathrm{d}\tilde{y}_2=\int_0^\infty\Phi\left(\frac{\tilde{y}_2}{\sqrt{y_1}}\right)f_{Y_1}(y_1)\mathrm{d}y_1, \tag{30}$$

where $\phi(\cdot)$ and $\Phi(\cdot)$ denote the PDF and CDF of standard Gaussian variable, respectively. Notice that $\Phi\left(\frac{\tilde{y}_2}{\sqrt{y_1}}\right)$ is piecewise monotonic with respect to $y_1$. Namely, for $\tilde{y}_2<0$, the value of $\Phi\left(\frac{\tilde{y}_2}{\sqrt{y_1}}\right)$ increase with the increasing of $y_1$; for $\tilde{y}_2>0$, $\Phi\left(\frac{\tilde{y}_2}{\sqrt{y_1}}\right)$ decrease with the increasing of $y_1$. In this regard, the analytical solution of the upper and lower bounds of the CDF of $Y_2$ can be derived as

$$\bar{F}_{Y_2}(y_2)=\begin{cases}\int_0^{\infty}\Phi\left(\sqrt{\dfrac{2\zeta\omega^3}{\pi S_0}}y_2\right)\bar{f}_{\Omega}(\omega)\mathrm{d}\omega, & y_2<0\\ 1/2, & y_2=0\\ \int_0^{\infty}\Phi\left(\sqrt{\dfrac{2\zeta\omega^3}{\pi S_0}}y_2\right)\underline{f}_{\Omega}(\omega)\mathrm{d}\omega, & y_2>0\end{cases}, \tag{31}$$

and

$$\underline{F}_{Y_2}(y_2)=\begin{cases}\int_0^{\infty}\Phi\left(\sqrt{\dfrac{2\zeta\omega^3}{\pi S_0}}y_2\right)\underline{f}_{\Omega}(\omega)\mathrm{d}\omega, & y_2<0\\ 1/2, & y_2=0\\ \int_0^{\infty}\Phi\left(\sqrt{\dfrac{2\zeta\omega^3}{\pi S_0}}y_2\right)\bar{f}_{\Omega}(\omega)\mathrm{d}\omega, & y_2>0\end{cases}, \tag{32}$$

respectively.

The analytical solutions [Eqs. (28), (31), and (32)] of $Y_1$ and $Y_2$ are derived directly based on the framework proposed in Section 2 without any numerical treatment. Only the upper and lower bounds of the natural frequency $\omega$ is employed, and thus illustrate the effectiveness of the proposed augmented random space scheme. These analytical solutions can be utilized to validate the numerical solution by the decoupled M-PDEM. In the numerical example, take $S_0=1$, $\zeta$=0.05, and

$$\bar{F}_{\Omega}(\omega)=\begin{cases}\Phi\left(\dfrac{\omega-\mu_1}{\sigma_2}\right), & \omega\le\mu_1\\ \Phi\left(\dfrac{\omega-\mu_1}{\sigma_1}\right), & \omega>\mu_1\end{cases}, \quad \underline{F}_{\Omega}(\omega)=\begin{cases}\Phi\left(\dfrac{\omega-\mu_2}{\sigma_1}\right), & \omega\le\mu_2\\ \Phi\left(\dfrac{\omega-\mu_2}{\sigma_2}\right), & \omega>\mu_2\end{cases}, \tag{33}$$

where $\mu_1=1.9$, $\mu_2=2.1$, $\sigma_1=0.1$, and $\sigma_2=0.2$.

Fig. 1 shows the comparison of the analytical solution of $Y_1$ given by Eq. (28) and the corresponding numerical solution obtained by the decoupled M-PDEM based on 200, 400, and 800 samples. It can be seen that the decoupled M-PDEM solution converge quickly to the analytical on with the increasing number of samples, and can reach high very high accuracy with the merely 800 samples used.

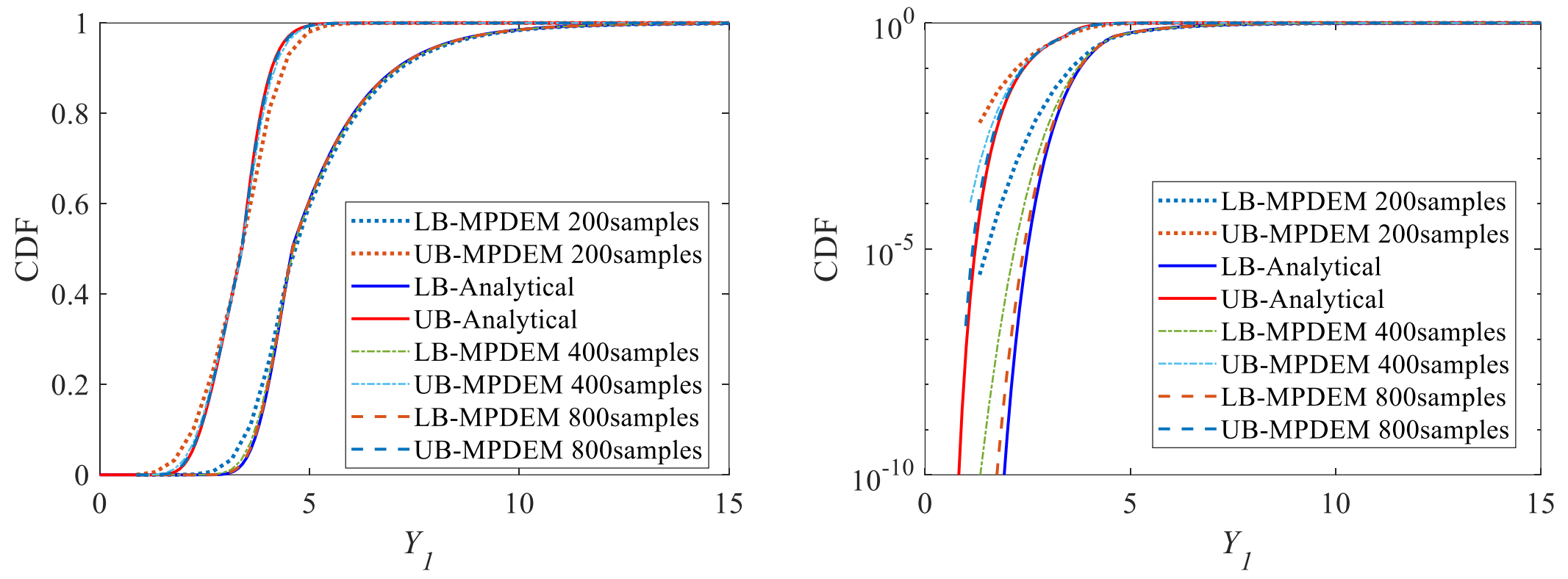

(a) linear coordinate (b) logarithmic coordinate

Fig. 1 Upper bounds (UB) and lower bounds (LB) of the CDF of $Y_1$ obtained by analytical solution and decoupled M-PDEM with various sample sizes.

Fig. 2 juxtaposes the analytical solutions and decoupled M-PDEM solutions for $Y_2$, which shows great agreement. Notice that in the numerical simulation of the decoupled M-PDEM, only $\omega$ is sampled, and the analytical conditional PDF given by Eq. (29) is substituted into the **step 3** of the numerical procedure of decoupled M-PDEM as the solution of Li-Chen equation for $X(t)$. This treatment naturally inserts analytically known information into the numerical implementation of decoupled M-PDEM, and greatly improves the accuracy. That is why, in this case, with only 200 samples, the decoupled M-PDEM solution can show such high accuracy.

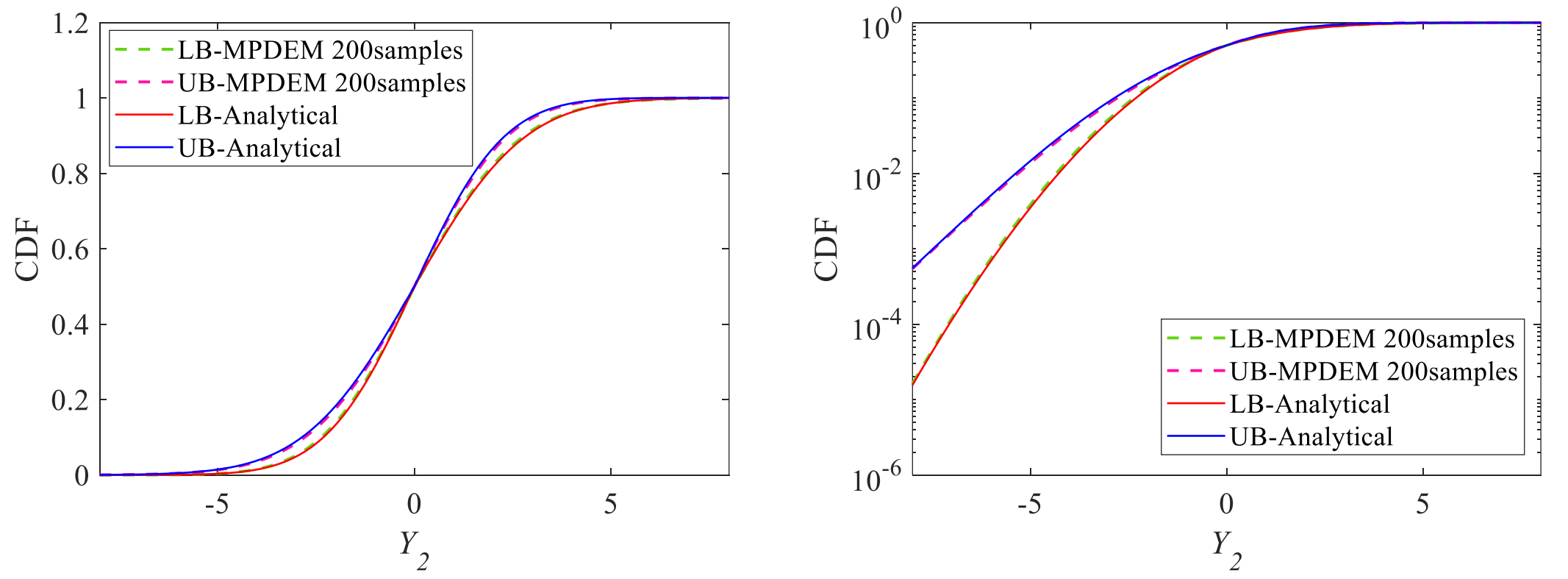

(a) linear coordinate (b) logarithmic coordinate

Fig. 2 Upper bounds (UB) and lower bounds (LB) of the CDF of $Y_2$ obtained by analytical solution and decoupled M-PDEM

## 4.2 10-DOF Bouc-Wen oscillator subject to imprecise excitation

This example demonstrates the application of the proposed framework to systems subject to excitations characterized by imprecise power spectral density (IPSD). Such problems are typically complex, necessitating a double-loop procedure; however, the proposed framework effectively decouples this complexity.

Consider a 10-DOF system with Bouc-Wen hysteretic model. The equation of motion is given by

$$\mathbf{M}(\boldsymbol{\Theta})\ddot{\boldsymbol{X}}(t)+\mathbf{C}(\boldsymbol{\Theta})\dot{\boldsymbol{X}}(t)+\boldsymbol{f}\left[\boldsymbol{X}(t),\boldsymbol{Z}(t),\boldsymbol{\Theta}\right]=\boldsymbol{B}\xi_I(t). \tag{34}$$

where $\boldsymbol{X}(t)$, $\dot{\boldsymbol{X}}(t)$, and $\ddot{\boldsymbol{X}}(t)$, respectively, denotes the displacement, velocity, and acceleration; $\boldsymbol{\Theta}=[\theta_1,\theta_2,\cdots,\theta_{20}]^{\mathrm{T}}$ is the vector of random parameters; $\mathbf{M}=\mathrm{diag}[m_1\theta_1,m_2\theta_1,\cdots,m_{10}\theta_{10}]$ is the diagonal mass matrix; $\mathbf{C}=a_0\mathbf{M}+a_1\mathbf{K}$ is the linear damping matrix in which

$$\mathbf{K}=\begin{bmatrix} k_1\theta_{11}+k_2\theta_{12} & -k_2\theta_{12} & & & \\ -k_2\theta_{12} & k_2\theta_{12}+k_3\theta_{13} & -k_3\theta_{13} & & \\ & \ddots & \ddots & \ddots & \\ & & -k_9\theta_{19} & k_9\theta_{19}+k_{10}\theta_{20} & -k_{10}\theta_{20} \\ & & & -k_{10}\theta_{20} & k_{10}\theta_{20} \end{bmatrix}, \tag{35}$$

and

$$\begin{pmatrix} a_0 \\ a_1 \end{pmatrix}=\frac{2\zeta}{\omega_1+\omega_3}\begin{Bmatrix} \omega_1\omega_3 \\ 1 \end{Bmatrix}; \tag{36}$$

$\omega_1$ and $\omega_3$ are respectively the square root of the first and third minimum eigenvalue of $\mathbf{M}^{-1}\mathbf{K}$; and $\zeta$ denotes the damping ratio. $\boldsymbol{Z}(t)=[Z_1,Z_2,\cdots,Z_{10}]^{\mathrm{T}}$ denotes the vector of the inter-story hysteretic components, and the restoring force is given by

$$\boldsymbol{f}\left[\boldsymbol{X}(t),\boldsymbol{Z}(t)\right]=\left[f_1^*-f_2^*,f_2^*-f_3^*,\cdots,f_{n-1}^*-f_n^*,f_n^*\right]^{\mathrm{T}}, \tag{37}$$

and

$$f_j^*\left[\boldsymbol{X}(t),\boldsymbol{Z}(t)\right]=k_j\left[\varepsilon X_j^*+(1-\varepsilon)Z_j\right], j=1,...,n, \tag{38}$$

in which $\boldsymbol{X}^*=[X_1,X_2-X_1,\cdots,X_n-X_{n-1}]^T$ denotes the vector of inter-story drift.

The evolution of the inter-story hysteretic component $Z_j$ satisfies (Ma et al. 2004)

$$\begin{aligned}\dot{Z}_j(t) &= g\left(\dot{X}_j^*, Z_j, \varepsilon_j\right) \\ &= \frac{h\left(\dot{X}_j^*, Z_j, \varepsilon_j\right)}{1+\delta_\eta \varepsilon_j}\left[A\dot{X}_j^* - \left(1+\delta_\nu \varepsilon_j\right)\left(\beta\left|\dot{X}_j^*\right|Z_j + \gamma \dot{X}_j^*\left|Z_j\right|\right)\right]\end{aligned} \quad (39)$$

In this equation, $\delta_\eta$ is the stiffness degradation parameter; $\delta_\nu$ is the strength degradation parameter; $A$, $\beta$, and $\gamma$ are the parameters for basic hysteresis shape control; $\varepsilon_j$ is proportional to the dissipated energy through hysteresis, which is given by

$$\varepsilon_j(t) = \int_0^t \dot{X}_j^*(\tau) Z_j(\tau)\, \mathrm{d}\tau . \quad (40)$$

In Eq. (39), $h\left(\dot{X}_j^*, Z_j, \varepsilon_j\right)$ is a pinching shape function, which takes the form

$$h\left(\dot{X}_j^*, Z_j, \varepsilon_j\right) = 1 - \zeta_1 \exp\left\{-\left[Z_j \operatorname{sgn}\left(\dot{X}_j^*\right) - qZ_{j,u}\right]^2 \Big/ \zeta_2^2\right\}, \quad (41)$$

where $\operatorname{sgn}(\cdot)$ is the signum function; $q$ is the pinching initiation parameter; $Z_{j,u}$ is the ultimate value of $Z_j$ given by

$$Z_{j,u} = \frac{A}{\left(1+\delta_\nu \varepsilon_j\right)\left(\beta+\gamma\right)}; \quad (42)$$

$$\zeta_1\left(\varepsilon_j\right) = \zeta_s\left[1 - e^{\left(-p\varepsilon_j\right)}\right]; \quad (43)$$

and

$$\zeta_2\left(\varepsilon_j\right) = \left(\psi + \delta_\psi \varepsilon_j\right)\left(\lambda + \zeta_1\right), \quad (44)$$

where $\zeta_s$ is the measure of total slip; $p$ is the parameter of pinching slope; $\psi$ represents the pinching magnitude; $\delta_\psi$ denotes the pinching rate; $\lambda$ is the parameter of pinching severity/rate interaction.

In Eq. (34), $\xi_I(t)$ is the excitation modeled by the IPSD function (Genovese & Sofi 2023) as

$$S(\omega)=\sigma^2\beta_0^I\left(\frac{\omega^2}{\omega^2+\left(\omega_H^I\right)^2}\right)\left(\frac{\left(\omega_L^I\right)^2}{\omega^4+\left(\omega_L^I\right)^4}\right)\frac{\rho_0}{\pi}\left(\frac{1}{\rho_0^2+\left(\omega+\Omega_0^I\right)^2}+\frac{1}{\rho_0^2+\left(\omega-\Omega_0^I\right)^2}\right), \quad (45)$$

where $\beta_0^I$ is defined so that the interval stochastic process $\xi_I(t)$ has variance $\sigma^2$; $\omega_L^I=\Omega_0^I+0.8\rho_0$, $\omega_H^I=0.1\Omega_0^I$, and $\Omega_0^I$ is the interval predominant circular frequency, $\rho_0$ is the frequency bandwidth.

In this numerical example, $[m_1,m_2,\cdots,m_{10}]=[2.6,2.6,\cdots,2.6]$ $(\times10^5\ \mathrm{kg})$, $[k_1,k_2,\cdots,k_{10}]=[4.9,2,2,2,1.8,1.8,1.8,1,1,1]$ $(\times10^7\ \mathrm{N/m})$, the damping ratio $\zeta=0.05$, and $\mathbf{B}=[1,1,\cdots,1]^\mathrm{T}$ $(\mathrm{kg})$. The other parameters associated with the restoring force [Eq. (38)] and the Bouc-Wen model are given in Table 1. $\theta_1,\theta_2,\cdots,\theta_{20}$ are independently distributed, in which $\theta_1,\theta_2,\cdots,\theta_{10}$ follow the uniform distribution $U(0.7,1.3)$ and $\theta_{11},\theta_{12},\cdots,\theta_{20}$ follow the lognormal distribution with mean equals to 1 and coefficient of variation equals to 0.15. Parameters for the interval excitation are taken as $\sigma^2=1.06(\mathrm{m^2/s^4})$, $\rho_0=17.33(\mathrm{rad/s})$, and $\Omega_0$ is bounded within $[23.91,45.22]$. The response quantity of interest is the maximum displacement of the first DOF, i.e.,

$$X_{\max}=\max\left(\left|X_1(t)\right|\right),\ \text{for}\ t\in[0,T], \quad (46)$$

where $T=10$.

Table 1. Values of parameters of Bouc-Wen model

| Parameter | $\varepsilon$ | $A$ | $\beta$ | $\gamma$ | $\delta_\nu$ | $\delta_\eta$ |
|---|---|---|---|---|---|---|
| Value | 0.04 | 1 | 15 | 150 | 1000 | 1000 |
| Parameter | $q$ | $p$ | $\delta_\psi$ | $\lambda$ | $\zeta_s$ | $\psi$ |
| Value | 0.25 | 1000 | 5 | 0.5 | 0.99 | 0.05 |

The double-loop MCS (DL-MCS) is employed to validate the efficiency and accuracy of the proposed method. Notably, the problem involves only one independent epistemic parameter $\Omega_0$. Therefore, in the outer loop of DL-MCS, 201 uniformly distributed points are selected to perform an exhaustive search across the interval $[23.91, 45.22]$. Within each inner loop, 10,000 samples are utilized for propagating the aleatory uncertainty. In both the DL-MCS and the proposed decoupled M-PDEM scheme, the excitations are generated using the spectral representation method (Shinozuka 1971). Consequently, additional random phase angles are introduced, significantly raising the overall dimension of the random variables involved to over 1,000. Thus, in the decoupled M-PDEM solution, the MCS samples are directly applied, with each sample assigned an equal probability as $1/N$.

Fig. 3 compares the results obtained by DL-MCS and the proposed decoupled M-PDEM scheme, with 2,000 samples utilized in the decoupled M-PDEM solution. Although a satisfactory agreement between the two results is observed, the accuracy of the decoupled M-PDEM is somewhat reduced compared to the previous example, despite the high sample size. This reduction likely arises from the significantly increased nonlinearity of the current problem, and also the much higher dimensionality of the random parameters involved, which hinders the precise determination of the probabilities assigned to each sample. Nevertheless, the decoupled M-PDEM scheme demonstrates substantially improved computational efficiency compared to the DL-MCS.

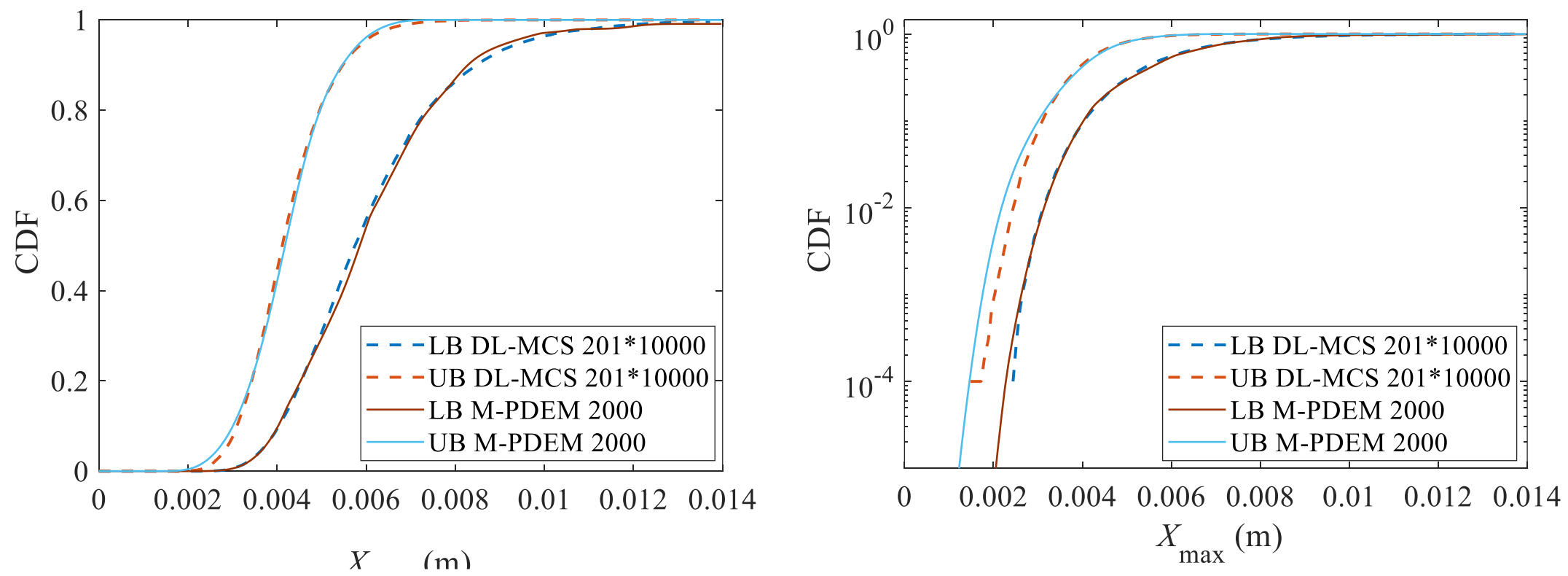

(a) linear coordinate (b) logarithmic coordinate

Fig. 3 Upper bounds (UB) and lower bounds (LB) of the CDF of $X_{max}$ obtained by DL-MCS and the proposed decoupled M-PDEM solution

## 4.3 Crash-box

Consider a crash box example as illustrated in Fig. 4, which simulates the process of a moving planar impacts on the crash box component. The output of interest is the final internal energy. As a representative benchmark case, the detailed model information is available in Reid (1998), and the analyses are conducted using the CAE Simulation Data Management Systems SCALE.sdm (https://www.scale.eu/en/products/scale-sdm/).

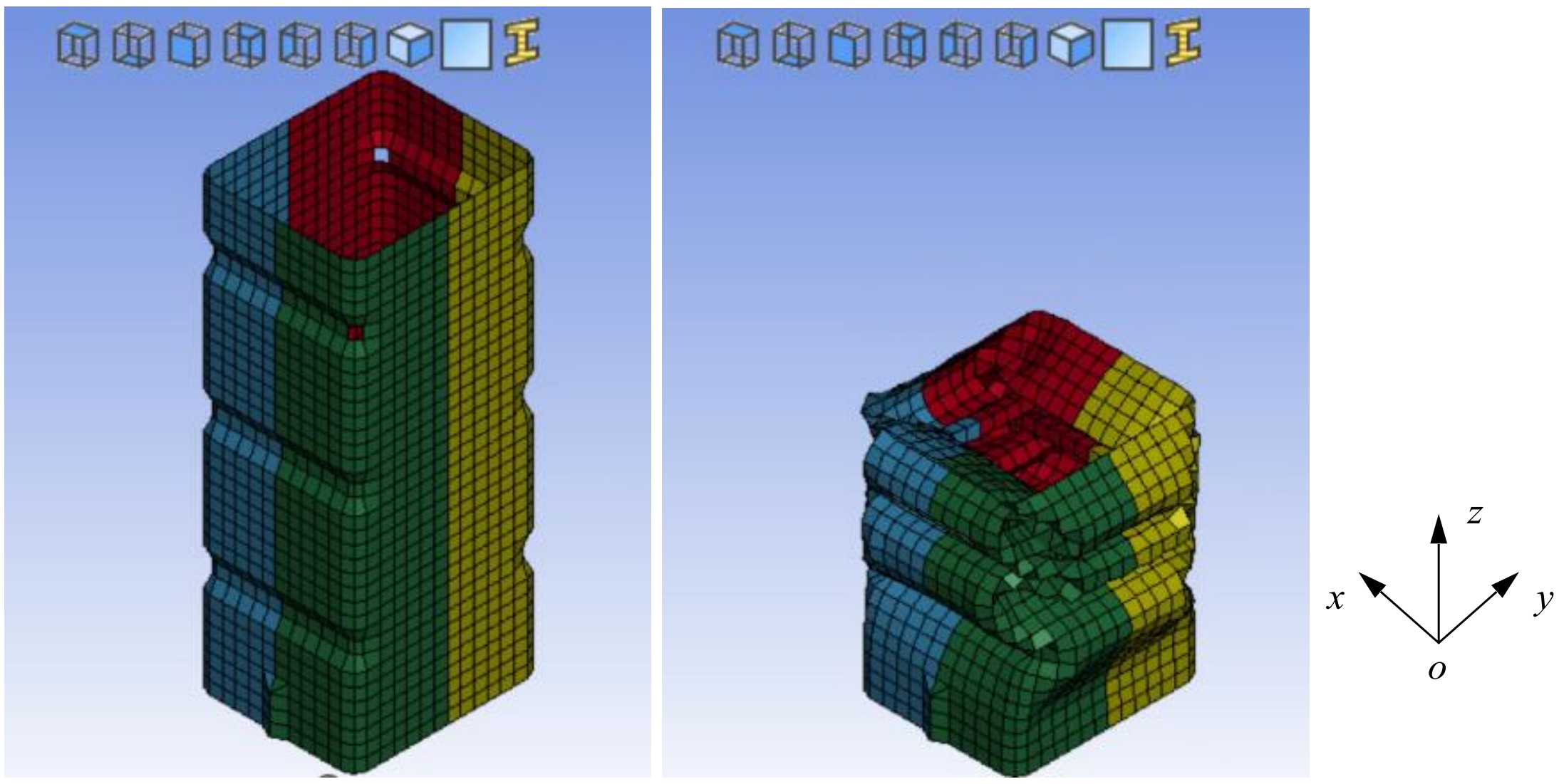

(a) before deformation (b) after deformation

Fig. 4 Schematic diagram of the crash box

The model includes five random input variables and two interval input variables, as listed in Table 2, with parameter values referenced from Hunkeler et al. (2013). The mass and velocity of the impactor were treated as primary epistemic inputs, and thus the assigned uncertainties to these two parameters are expanded to consider a broader range of load cases.

Table 2. Uncertainty input parameters for crash box example

| Variable | Description | Uncertainty model |
|---|---|---|
| $S_X$ | dimensional scale factor in $x$ direction | $N(1,1/60)$ |
| $S_Y$ | dimensional scale factor in $y$ direction | $N(1,1/60)$ |
| $\tau$ | scale factor of the thickness of the tube | $N(1,0.03286)$ |
| $\alpha_X$ | offset of the attack angle in $x$ direction | $U(-1°,1°)$ |
| $\alpha_Y$ | offset of the attack angle in $y$ direction | $U(-1°,1°)$ |
| $M_I$ | mass of the planar impactor | [-650,950] (kg) |
| $v_I$ | velocity of the planar impactor | [7,11] (m/s) |

$N(\mu,\sigma)$ denotes Gaussian distribution with mean equals to $\mu$, and standard deviation equals to $\sigma$; $U(a,b)$ denotes uniform distribution between $a$ and $b$; and $[a,b]$ denotes the interval variable with the given bounds.

Due to the prohibitive computational cost of this model, the DL-MCS approach is infeasible. Moreover, given the relatively clear monotonic relationship, where larger $M_I$ and $v_I$ typically lead to higher internal energy, the Vertex MCS method is adopted for validation. This method evaluates the maximum and minimum values of both interval inputs, yielding a total of $2\times1000=2000$ samples.

Fig. 5 compares the p-box boundaries of the output obtained via the proposed decoupled M-PDEM scheme (with 200 representative analyses) and the Vertex MCS results. Since only 1,000 samples are used for Vertex MCS, the results on logarithmic coordinate are not shown. The proposed solution successfully envelops the Vertex MCS bounds, though the latter produces slightly tighter boundaries. This confirms the reliability of the decoupled M-PDEM solution, as the Vertex MCS results are always equal to or tighter than the true CDF bounds. To further verify the accuracy, the conditional CDFs obtained by the decoupled M-PDEM at the interval vertices are also plotted, showing close agreement with the Vertex MCS results. This demonstrates the accuracy of the proposed procedure.

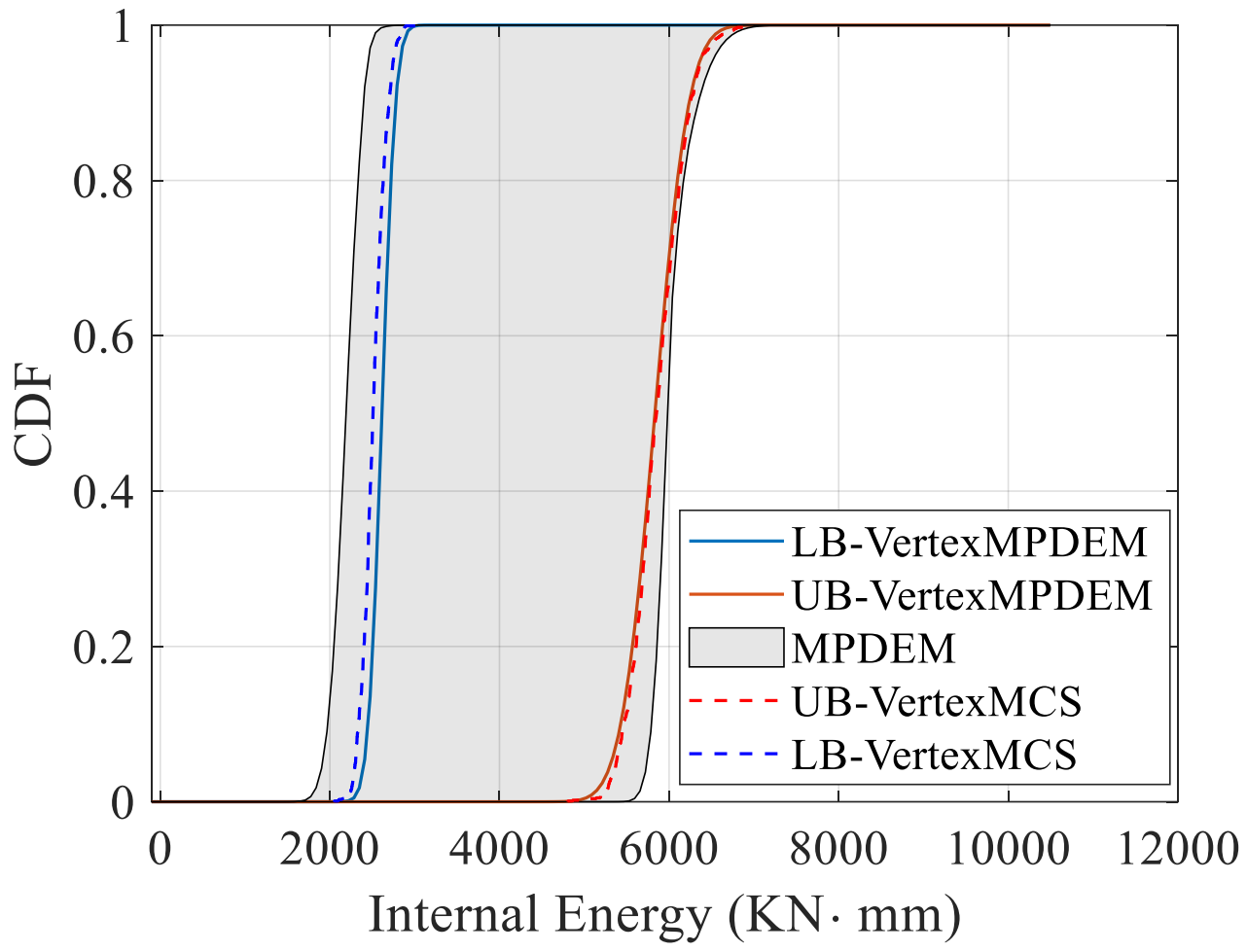

Fig. 5 Comparison of the decoupled M-PDEM results with Vertex MCS results of the crash box example

# 5 Concluding Remarks

In this work, a unified framework for hybrid aleatory-epistemic uncertainty propagation in dynamical systems has been developed based on the decoupled multi-probability density evolution method (M-PDEM). The proposed framework leverages an augmented random space representation, allowing for the decoupled treatment of various forms of epistemic uncertainty including distribution-free p-boxes. This transforms the hybrid uncertainty propagation problem into a unified task of solving conditional probability distributions, which is efficiently addressed using the decoupled M-PDEM. The framework's effectiveness and versatility have been demonstrated through various numerical examples involving diverse types of uncertain inputs. The proposed approach not only provides an efficient tool for the propagation of hybrid uncertainty in complex dynamical systems, but also constitutes a significant extension of the PDEM methodology.

Despite these advancements, current results highlight opportunities for further improvement, particularly in enhancing the accuracy of the decoupled M-PDEM solution. First, while the approach performs efficiently with one or two epistemic parameters, its extension to higher-dimensional epistemic uncertainty remains challenging due to the limited number of samples affordable. Second, in reliability analysis where rare events are critical, refinements are needed to capture tail behavior accurately. These limitations could be effectively addressed by integrating advanced sampling techniques and surrogate models in future developments.

## Acknowledgement

The authors are grateful for the support by the European Union's Horizon 2020 research and innovation programme under Marie Sklodowska-Curie project GREYDIENT – Grant Agreement n◦955393, the German Research Foundation (DFG) (Grant No. FR 4442/2-1) and the National Natural Science Foundation of China (Grant No. 12302037). The authors also highly appreciate SCALE GmbH team for their support in the completion of the crash box example. The discussions and suggestions of Prof. Jian-Bing Chen from Tongji University is highly appreciated.